\documentstyle[12pt]{article}
\input{epsf}
\begin{document}
\title{On the asymptotic behavior of the super sonic interstellar 
gas flow which is made by spiral density wave, propagating in rapidly 
rotating galaxy disk}
\author{E.V. ~Liverts \\ 
Dept. of The Theoretical Physics,\\
Novgorod State University,\\
Novgorod, 173003, Russia}
\date{${July ~6th,~1998}$}
\maketitle
\begin{abstract}
Nonlinear solutions for the large-scale flow of interstellar gas
in the presence of a spiral gravitational field have received 
considerable attention because this model allows to explain a 
lot of observations for galaxy disks. However this investigations
were forced to limit by numerable analysis of the problem
because of its nonlinearity.

In this paper we wish to carry out the analytical expression which
allows us to describe the super sonic nonlinear interstellar gas flow
in rapidly rotating galaxy disk which is made by the spiral density
wave.

One of characteristic parameters of theory is the amplitude of spiral 
density wave potential corresponding to separatrix, which separates 
super sonic flows from flows containing the jump from super sonic flows 
to subsonic.    

We have defined the dependence of perturbing potential
value, under which the galaxy shock wave appeared with respect to
the parameters characterizing the gas disk (such as sound speed in gas,
disk rotation speed, "spiral design" rotation speed).
\end{abstract}

In recent years there have been several numerical simulations for 
nonlinear solutions of large scale flow of interstellar gas in 
the stellar gravitational field of spiral galaxies (Lin, Yuan \& 
Shu 1969; Roberts 1969; Woodward 1975) by using the wave theory
of spiral structure of galaxy. This concept has played a remarkable 
role in understanding of many processes occurring in the galaxies 
such as the structure of dust lanes, observed along the inner edges 
of spiral arms in many galaxies (Fujimoto, 1966; Lynds, 1970), the 
enhanced synchrotron radiation from spiral arms (Mathewson 
{\it et al.,} 1972; Sofue, 1985), the radio emission of HI at 
the wavelength \mbox{$\lambda=21 ám$} (Berman \& Mishurov 1980), 
a trigger mechanism for star formation, creating the narrow bands 
of young highly luminous stars which delineate the spiral arms 
(Roberts, 1969; Shu {\it et al.}, 1972; Shu, 1973) and etc.
It was shown that the spiral density wave, propagated in the stellar
galaxy disk and interpreted as the galaxy spiral arms, is induced 
large nonlinear perturbations in the axial symmetric gas flow of the 
galactic disk. Such nonlinear phenomena take place in the gas even 
though the amplitude of the spiral stellar field is small. This is 
because the response to a spiral gravitational potential induced by 
the stellar density wave is roughly proportional to $a^{-2}$, where 
$a$ is the velocity dispersion, for stars, or sound speed for the gas 
(Woodward 1975). For the gas $a \sim 8~\frac{km}{s}$, while for the 
stars $a \sim 40~\frac{km}{s}$. So, if the amplitude of the perturbation 
field is a few per cent and we can use the liner theory to calculate 
a disturbance in the stellar disk, the perturbation of the gas disturbance 
is much stronger and we must use nonlinear theory. Nevetherless, the 
nonlinearity makes this model more contextual in the sense of physical 
ideas, which let to explain the morphology peculiarities of building 
and evolution of galaxies.

In this paper we wish to carry out the analytical expression which
allows us to describe the disturbance of the super sonic nonlinear 
interstellar gas flow in rapidly rotating galaxy disk which is made 
by the spiral stellar density wave.

The hydrodynamic equations for the steady nonlinear gas flow (self-similar
solutions) in the frame rotating with angular velocity $\Omega_p$ are,
at position ${\bf r}$,

\begin{equation}
\nabla\cdot ( \rho{\bf  v})=0
\end{equation} 

\begin{equation}
({\bf     v}\cdot\nabla){\bf    v}+2{\bf
\Omega}_{p}\times{\bf       v}=-\frac{1}{\rho}        \nabla
P-\nabla(\Phi-\frac{1}{2}{\Omega_{p}}^{2}r^{2})
\end{equation} 

where $\rho$, ${\bf v}$ and $P=a^2\rho$ (we have assumed that the gas is 
isotermal, i.e. $a$ is isothermal sound speed) are respectively the density, 
velocity, and pressure of the interstellar gas; $\Phi$ is the gravitational
potential of the stars contained an unperturbed axisymmetric part as well as
non axisymmetric perturbation induced by spiral wave; $r$ is a distance
from the centre of the galaxy disk. In the absence of a stellar density wave
the gas flow is supposed to be purely circular, so the gravitational 
potential $\Phi$ is axisymmetric and can be determined by the condition 
$\Phi(r)=\frac{1}{2}\Omega^2r^2$, where $\Omega=\Omega(r)$ is an angular 
velocity of rotating galaxy disk at position $r$

The unperturbed axisymmetric solutions of equations (1) and (2) for 
rapidly rotating thin disk are

$$
v_{z0}=0, \quad v_{r0}=0, \quad v_{\phi 0}=r(\Omega-\Omega_p)
$$

where, $ v_{r0}, v_{\phi 0}, v_{z0}$ are in cylindrical polar coordinates.

Describing the nonlinear gas flow perturbed by the spiral stellar density
wave each hydrodynamic variable can be written as a combination of an
unperturbed axisymmetric part and a perturbation, denoted by subscripts 0
and 1 respectively (Roberts 1973) then the gravitational potential 
is \mbox{$\Phi=\Phi_0+\Phi_1$}. According to other authors, beginning from Roberts 
(1973) it can be suggested that the perturbing potential $\Phi$ is similar 
to the spiral. It is then useful to introduce the spiral coordinates 
$(\eta,\xi,z)$. The spiral coordinates is the set of coordinates fixed in 
the rotating frame that are parallel and perpendicular to the spiral 
equipotential square so $\eta$ is constant along the spiral arms while 
$\xi$ is constant everywhere on lines orthogonal to the spiral arms and 
$z$ is constant everywhere on plane parallel to galaxy disk. This 
coordinates are related to the plane cylindrical polar coordinates 
$(r, \phi)$ by 
$$\eta=-\frac{2}{\tan i}\ln r+2\phi$$
$$\xi=-2\ln r+\frac{2}{\tan i}\phi$$
where $i$ is angle of inclination of a spiral arm to the circumferential 
direction; this angle is related to the radial wavenumber of the spiral 
wave $k$ by conditions $\tan i=-\frac{2}{kr}$, and $k=-\frac{2 \pi}
{\lambda}$; $\lambda$ is the wave length, so we get 

\begin{eqnarray*}
d\eta=-k dr+2d \phi \\
d\xi =-2\frac{d r}{r}-krd \phi
\end{eqnarray*}

The unperturbed gas velocity components parallel to the $(\xi,\eta, z)$
directions respectively are

$$v_{\| 0}=(\Omega - \Omega_{p})r\cos i, 
v_{\bot 0}=(\Omega - \Omega_{p})r\sin i=\frac{2}{k}(\Omega-\Omega_p)\cos i,$$

where $v_{\|},v_{\bot}$ are the gas velocity components parallel and 
perpendicular to the spiral arms respectively. It is easy to see that
$\frac{v_{\bot 0}}{v_{\| 0}}\sim\frac{1}{kr}$.
Since each hydrodynamic variable can be written as a combination
of an unperturbed part and a pertubation we can define as 
$v_\|=v_{\|0}+v_{\|1}$,\quad $v_\bot=v_{\bot 0}+v_{\bot1}$,\quad
$\rho=\rho_0+\rho_1$

According to Roberts (1969) we can assume that the spirals are tightly 
wound, i.e. $i \ll 1$, it corresponds to the short, comparing with scale
$r$, waves, i.e. $\lambda \ll r$, and allows to use WKB approximation 
for the spiral wave description. We would like to note that the spirals 
in this case appear to be "tightly" wound.
Then it can be shown (Nelson \& Matsuda 1977), that the gradients for
the spiral perturbations satisfy

\begin{eqnarray*} 
\frac{\partial}{\partial \xi} ~\leq ~\sin^2~i~\frac{\partial}{\partial \eta},
\end{eqnarray*} 

and it can be also found that for any function $\phi (r)$ with scale length 
$r$ i.e. $\frac{d\phi}{dr}\sim 1$ we get

\begin{eqnarray*} 
\frac{\partial}{\partial \eta} ~\sim ~\sin~i~\frac{\partial}{\partial \eta}
\log(v_{\|1},v_{\bot1},\rho_1),
\end{eqnarray*} 

In this approximations it can be neglected derivatives of perturbed 
quantities in the direction of the "slow" coordinate $\xi$ along the 
spiral arms with respect to their derivatives in the direction of 
the "rapid" coordinate $\eta$ which is perpendicular to the spiral arms. 
In addition, it can be said that the typical length in this problem is 
$\frac{1}{k}\sim \lambda$, besides that $\frac{\lambda}{r},\frac{\Delta}
{\lambda}\ll 1$ (here $\Delta$ is a typical thickness of the galaxy disk 
at position $r\sim 10kpc$) than because of equation (1) $v_z=0$ (it is 
right because in such conditions the last term $\frac{\partial}{\partial z}$.
in equation (1) is dominant one)
Finally assuming the perturbation potential as
\begin{eqnarray*} 
\Phi_1(\eta)=F\frac{\Omega^2 r \cos^2i}{k}\cos \eta
\end{eqnarray*} 
where $F$ is the amplitude of the perturbation potential due to the 
stellar density spiral wave as a fraction of the unperturbed axisymmetric
gravitational potential.

It can be introduced scale \mbox{$\frac{\chi^2}{2\Omega}\frac{\cos i}{k}=
11.34\frac{km}{s}$} for $v_\|$ as well as scale for $v_\bot$ 
\mbox{$\chi\frac{\cos i}{k}=18.1\frac{km}{s}$} than \mbox{$v_{\bot 0}
\frac{k}{\chi\cos i}=2\frac{\Omega-\Omega_p}{\chi}=-\nu=0.722$}
Here are used the same set of parameters as used by Shu {\it et al} 
(1973): \mbox{$\Omega=24.7~\frac{km}{s~kpc}$}; \mbox{$\Omega_p=13.5~
\frac{km}{s~kpc}$}; \mbox{at position $r=10~kpc$ };
\mbox{$\chi=2\Omega((1+\frac{r}{2\Omega}
\frac{d \Omega}{d r}\cos^2 i))^\frac{1}{2}=31.0~\frac{km}{s~kpc}$} 

Eliminating $\rho_1$ between equations (1) and (2), according to Shu 
{\it et al} (1973) we can write this equations in dimensionless form 
\begin{eqnarray}
\frac{(-\nu+u)^2-c}{-\nu+u}~\frac{du}{d\eta}=v-f\sin~\eta \nonumber \\
u+(-\nu+u)\frac{dv}{d\eta}=0
\end{eqnarray} 
where $u,~v$ are the dimensionless velocity variables perpendicular 
$v_\bot$ and parallel $v_\|$ to the spiral arms respectively; 
$f=F\frac{\Omega^2}{\chi^2}kr\sim 0.1-0.2$ is the dimensionless 
amplitude of the stellar density spiral wave; $c=\frac{a^2}{\chi^2
\frac{\cos^2 i}{k^2}}=0.195$ is the square of the dimesionless sound speed

Equations (3) have been solved numerically by other authors mentioned
above. Eliminating $v$ between these equations we can get a equation 
which is to be solved for $u$
\begin{equation}
\frac{(-\nu+u)^2-c}{-\nu+u}~\frac{d^2u}{d\eta^2}+\frac{(-\nu+u)^2+c}
{(-\nu+u)^2}~(\frac{du}{d\eta})^2+\frac{u}{-\nu+u}=-f\cos~\eta
\end{equation}ÿ

Beginning with some value of $f$ the solution of equation (4) contains
an irregularity is due behavior of the equation solution in neighborhood
some point at $u=\nu+\sqrt{c}$. If we introduce a new 
variable by condition $\tilde u=u-\nu-\sqrt{c}$ then we get

$$
\frac{(2\sqrt{c}+\tilde u)}{2(\sqrt{c}+\tilde u)}\tilde u\tilde u''+
\Bigl(1-\frac{2\sqrt{c}+\tilde u}{2(\sqrt{c}+\tilde u)^2}\tilde u\Bigr)
\tilde u'^2-\frac{\nu}{2\sqrt{c}(\sqrt{c}+\tilde u)}\tilde u=-\frac{1}
{2}(1+\frac{\nu}{\sqrt{c}}+f\cos\eta)
$$

In the flow zone, which lies in the neighborhood of this point, gas 
parameters corresponding to the explosive solution are exposed by sharp 
changes, and along the spatial coordinate a narrow transitional area 
appears, in this area gas parameters constantly change: a so called jump 
of density if in the chosen frame this area is stable, or a shock 
wave if the transitional area is transferred in space with respect to time.
Nevertheless, for the counting of flow parameters in this area we need
to take into account the additional non-gas dynamic transfer of impulse 
and energy, which corresponds to the viscosity and heat of dissipation. 
The latter may be taken into account if we add in the simplest
case "viscous pressure" to the gas-dynamic pressure.

By assuming in (4) that $u=f\nu y$ and $\omega^2=(\nu^2-c)^{-1}$ we get
\begin{eqnarray*}
y''+\omega^2 y&=&\omega^2\cos\eta+f((2\nu^2\omega^2+1)yy''+
(2\nu^2\omega^2-1)y'^2+ \\
& + &\omega^2(y^2-2y\cos\eta))+f^2\nu^2\omega^2(-3y^2y''-2yy'^2+
y^2\cos\eta)+ \\
& + & f^3\nu^2\omega^2(y^3y''+y^2y'^2)
\end{eqnarray*}
If we choose the terms in the same powers of $f$ we get the following set
of equations which are to be solved successively for $y_1; y_2; y_3; \dots $

\begin{eqnarray*}
{y_0}''+ \omega^2 y_0&=&\omega^2\cos\eta \\  \\
{y_1}''+ \omega^2 y_1&=&(2\nu^2\omega^2+1)y_0{y_0}''+
(2\nu^2\omega^2-1){y_0}'^2+\omega^2(y_0^2-2y_0\cos\eta) 
\end{eqnarray*}
\begin{eqnarray*}
{y_2}''+ \omega^2 y_2 &=& (2\nu^2\omega^2+1)(y_0{y_1}''+y_1{y_0}'')+
(2\nu^2\omega^2-1)2{y_0}'{y_1}'+ \\ 
&+&\omega^2(2y_0y_1-2y_1\cos\eta)+\nu^2\omega^2(-3y_0^2{y_0}''
-2y_0{{y_0}'}^2+y_0^2\cos\eta) \\
\end{eqnarray*}
\begin{eqnarray*}
{y_3}''+ \omega^2 y_3&=&(2\nu^2\omega^2+1)(y_0{y_2}''+y_1{y_1}''+
y_2{y_0}'')+ \\
&+&(2\nu^2\omega^2-1)({{y_1}'}^2+2{y_0}'{y_2}')+\omega^2(y_1^2+2y_0y_2-
2y_2\cos\eta)+ \\ 
&+&\nu^2\omega^2(-3(y_0^2{y_1}''+2y_0y_1{y_0}'')- 
2(2y_0{y_0}'{y_1}'+y_1{{y_0}'}^2)+ \\
&+&2y_0y_1\cos\eta)+\nu^2\omega^2(y_0^3{y_0}''+ y_0^2{{y_0}'}^2) 
\end{eqnarray*}
$\cdots \cdots \cdots \cdots \cdots \cdots \cdots \cdots \cdots$

This set of equations has been used by Shu,Milione \& Roberts (1973)
for discussing the slightly nonlinear regime and comparing it with the 
results of the linear theory by Lin {\it et al.} (1969). At the same 
time the authors noted that for such value as realistic gravitational 
potential, the series which are solutions of this equation set may 
converge very slowly, if at all. So they used a numerical method of 
investigation this problem and they got that for the typical value 
of parameter $f$, such as $f_c=0.105$ the behavior of the solution was 
changed and if $f$ was being increased above $f_c$ the solution could 
not remain smooth everywhere and had to contain a jump. Physically, 
the jump corresponds to a shock wave which are formed at the transition 
from supersonic gas flow to subsonic.

The solution $y(\eta)$ can be written in the form 

\begin{equation}
y(\eta)=\sum_{i=0}^{\infty}\sum_{k=0}^{\infty}A_{i,k}\cos(k\eta)  
\end{equation}

In summation over $i$ the dominant contribution arising from a first 
few terms because if $i$ increases the series coefficient $A_{i,k}$
decreases very quickly, on the contrary, in summation over $k$ dominant 
contribution is arisen from the remote terms of this series because 
the coefficient $A_{i,k}$ is changed with respect to $k$ very slowly. 
The necessity to take into account the remote terms is so important 
as the gas flow velocity is more closely to sound speed, this corresponds 
with that the value of parameter $f$ is approached to its typical 
value mentioned above. 

The main coefficient $A_{0,n}$ can be obtained from solution of $n-1$ 
equation, to do it we must join terms which are proportional to
\mbox{$\cos(n+1)\eta$}, if \mbox{$n > 0$} we get

\begin{eqnarray}
y_n=\frac{(n+1)^2}{(n+1)^2-\omega^2}(\nu^2\omega^2S_1(n)+\frac{1}{2}S_2(n)
-\frac{\omega^2}{(n+1)^2}(S_1(n)-a_{n-1})+ \nonumber\\
+\frac{\nu^2\omega^2}{4}(\frac{S_1(n-1)}{(n+1)^2}-3S_3(n)-2S_4(n))+
\frac{\nu^2\omega^2}{8}(S_5(n)+S_6(n)))\cos(n+1)\eta+ \nonumber\\
+\underbrace{\ldots\ldots\ldots\ldots\ldots\ldots\ldots}_{\mbox{the terms
contained $\cos((n-1)\eta)$ and etc.}}
\end{eqnarray}

Where

\begin{eqnarray*}
S_1(n)=\left\{
\begin{array}{ll}
\frac{1}{2}a_{\frac{n-1}{2}}^2+\sum_{k=0}^{\frac{n-3}{2}}a_k a_{n-1-k}&
\mbox{if $n$ is odd} \\ 
\bigskip
\sum_{k=0}^{\frac{n-2}{2}}a_k a_{n-1-k}&
\mbox{if $n$ is even} 
\end{array} \right.
\end{eqnarray*}

\bigskip
\begin{eqnarray*}
S_2(n)=\left\{
\begin{array}{ll}
\sum_{k=0}^{\frac{n-3}{2}}(\frac{n-2k-1}{n+1})^2a_k a_{n-1-k}&
\mbox{if $n$ is odd} \\
\bigskip
\sum_{k=0}^{\frac{n-2}{2}}(\frac{n-2k-1}{n+1})^2a_k a_{n-1-k}&
\mbox{if $n$ is even} 
\end{array} \right.
\end{eqnarray*}

\bigskip
\begin{eqnarray*}
S_3(n)=\sum_{k=0}^{n-2}\sum_{m=0}^{n-2-k}
(\frac{n-1-k-m}{n+1})^2a_ka_ma_{n-2-k-m}
\end{eqnarray*}

\bigskip
\begin{eqnarray*}
S_4(n)=(\sum_{k=0}^{n-2}\sum_{m=0}^{n-2-k}\frac{(k+1)(m+1)}{(n+1)^2}
a_ka_ma_{n-2-k-m}
\end{eqnarray*}

\bigskip
\begin{eqnarray*}
S_5(n)=\sum_{k=0}^{n-3}\sum_{m=0}^{n-3-k}\sum_{l=0}^{n-3-k-m}
(\frac{n-2-k-m-l}{n+1})^2a_ka_ma_la_{n-3-k-m-l}
\end{eqnarray*}

\bigskip
\begin{eqnarray*}
S_6(n)=\sum_{k=0}^{n-3}\sum_{m=0}^{n-3-k}\sum_{l=0}^{n-3-k-m}
\frac{(m+1)(n-2-k-m-l)}{(n+1)^2}a_ka_ma_la_{n-3-k-m-l}
\end{eqnarray*}

where $a_n$ is a coefficient at $\cos(n+1)\eta$ in the solution $y_n$
which is related to $A_{0,n}$ by relation $A_{0,n+1}=f^na_n$;
\mbox{$a_0=\frac{\omega^2}{\omega^2-1}$}
Recursion relation for this coefficients can be obtained by using the
successive approximations method to solve the equation (6) for $n$
approach.
The first approach can be written in form
\begin{equation}
a_n=\omega^2(1.89fa_0(\nu^2\omega^2+\frac{1}{2}))^n\prod_{m=0}^n
\frac{(m+1)^2}{(m+1)^2-\omega^2}
\end{equation}
For large $n$, taking into account that the product approximates rapidly,
it can be substituted by limit $\prod_{m=1}^\infty(1-\frac{\omega^2}{m^2}
)=\frac{\sin~\pi\omega}{\pi \omega}$ (equality $\omega$ to whole number 
corresponds to the Lindblad's resonance). So for series (5) the following
expression can be written: 
\begin{equation}
y\simeq \frac{\omega^2}{\omega^2-1}\cos(\eta)-0.0342\frac{\pi \omega^3}
{\sin~\pi \omega}\sum_{k=1}^\infty(1.89f(\nu^2\omega^2+\frac{1}{2})\frac
{\omega^2}{\omega^2-1})^k\cos~(k+1)\eta
\end{equation}
The numerical coefficients appeared in the last two expressions are the
results of averaging the difference between the product of final number 
of the terms $\prod_{m=0}^n\frac{(m+1)^2}{(m+1)^2-\omega^2}$ and its 
limit when the number is infinite and the result of the calculation of 
the sums $S_1(n),S_2(n)$ and etc. The latter can be done while using 
the method of continued fraction.

Trigonometric series in (8) in condition that \mbox{$\mid 1.89f(\nu^2
\omega^2+\frac{1}{2})\frac{\omega^2}{\omega^2-1}\mid < 1$} converges and,
according to Prudnikov, Brychkov \& Marichev (1981) can be expressed in
elementary functions, so we have:
\begin{equation}
y\simeq \frac{\omega^2}{\omega^2-1}\cos(\eta)-0.0342\frac{\pi \omega^3}
{\sin~\pi \omega}\frac{cos~\eta-\alpha}{1-2\alpha\cos~\eta+\alpha^2}
\end{equation}
Here 
$$
\alpha=1.89f(\nu^2\omega^2+\frac{1}{2})\frac{\omega^2}{\omega^2-1}
$$
Using (9) we will put down at last the expression which consists the
main part of speed component perpendicular to wave front, i.e. $u_\bot$,
taking into account that (9) to the large extend is adequate to large
$k$ from (5), we will change in (9) at least the first member by the 
exact solution of the equation for the first approximation, so we get:
\begin{eqnarray}
u_\bot=u_{\bot0}+u_{\bot1}=-\nu(1+\frac{f^2}{2}(\frac{\omega^2}
{\omega^2-1})^2-f\frac{\omega^2}{\omega^2-1}\cos~\eta \nonumber\\
+f^2\frac{\omega^2}{\omega^2-1}(\frac{\omega^2}{(\omega^2-1)
(\omega^2-4)}(2\nu^2\omega^2+\frac{\omega^2}{2}-1))\cos~2\eta \nonumber\\
+0.0342f\frac{\pi \omega^3}{\sin~\pi \omega}\frac{cos~\eta-\alpha}
{1-2\alpha\cos~\eta+\alpha^2})
\end{eqnarray}

Integral curves (10) of equation (4) for several values of parameter $f$
are given on Figure 1. As we can see from this picture with parameter $f$
increasing, i.e. with the growth of perturbed potential amplitude the
gas flow speed, at $\eta=0$, is decreasing and may reach sound speed in 
gas. The integral curve, corresponding to this condition, has a sharp jump
if $\eta=0$ and presents a separatrix which separates the solutions 
corresponding to the gas flow with supersonic speed from the solution,
in which there are areas of both supersonic and subsonic gas flow. 
If one prefers to see a results represented as the velocity plane its
pictured at Figure 2. 
\begin{figure}
\epsfxsize=15cm  \epsfbox{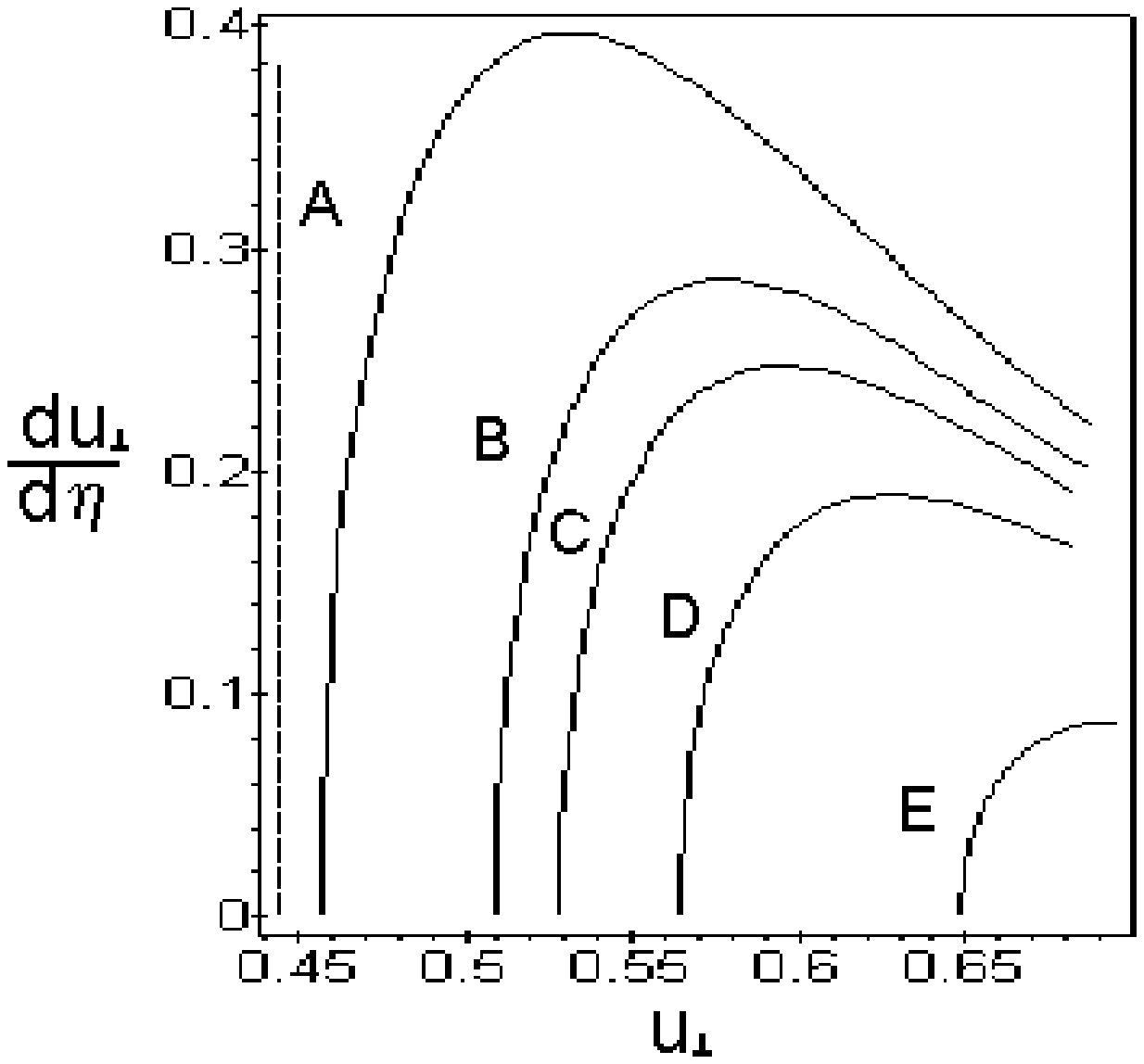}
{
Figure 1. The phase portrait of equation. Curve $A$ assumes the 
amplitude of the perturbation potential such as $f=0.11$; $B$ 
assumes $f=0.10$; $C$ assumes $f=0.095$; $D$ assumes $f=0.085$; 
$E$ assumes $f=0.055$; The sound speed is marked by broken line
}
\end{figure}

\bigskip
\begin{figure}
\epsfxsize=15cm  \epsfbox{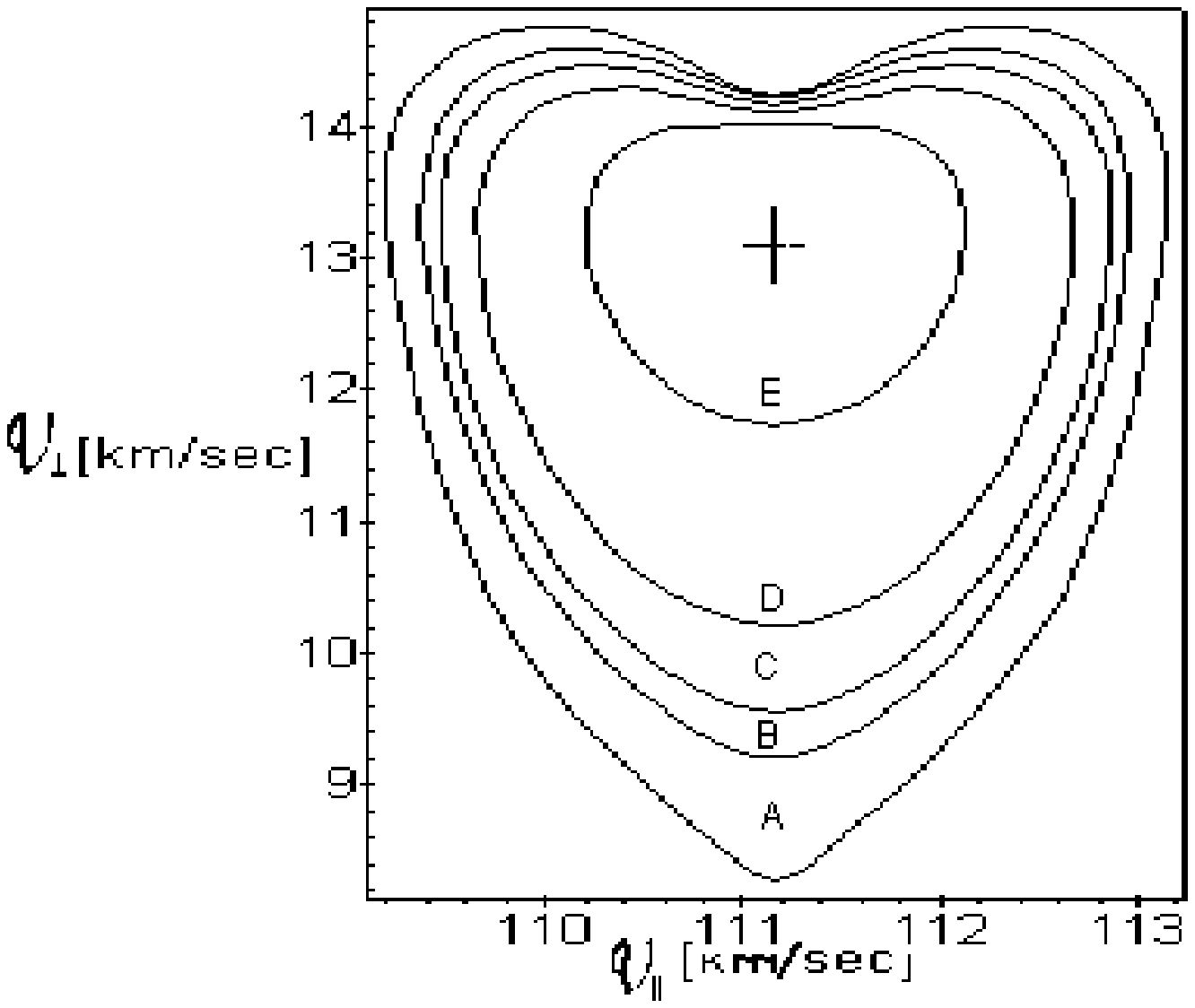}
{
Figure 2. The supersonic flow in the velocity plane. Curve $A$ assumes 
the amplitude of the perturbation potential such as $f=0.11$; $B$ 
assumes $f=0.10$;  $C$ assumes $f=0.095$; $D$ assumes $f=0.085$; 
$E$ assumes $f=0.055$;  For $f=0.11$ the flow contains a cusp at 
the location of the minimum of the spiral potential, $\eta=0$. 
The unperturbed axisymmetric velocity is marked by a cross.
}
\end{figure}

\bigskip

The amplitude value of perturbed potential $f_c$ which corresponds to this
separatrix was defined by Shu {\it et al.} numerable. Formula (10) 
allows to define the dependence $f_c$ respect to the parameters of the 
rotating gas disk. For this we take in (10) $u_\bot=\sqrt{c}$ if 
$~\eta=0$, then solving the equation in respect to $f$ we get:
$$
f_c\simeq -\frac{1}{2}\frac{\sin~\pi\omega}{\pi\omega^3}
\frac{\frac{\sqrt{c}}{\nu}+1}{0.0342-0.0616(1-1.89(1+\frac{\sqrt{c}}{\nu})
(\nu^2\omega^2+\frac{1}{2}))}\simeq 0.12
$$
Here the same parameters with which Shu {\it et al} (1973) solve
equation (3) numerable, and the value we get is practically coincides
with their result.

The author thanks Professor Yu.N.Efremov \& Professor A.D.Chernin from
Sternberg Astronomical Institute, Moscow State University for suggesting 
this problem as well as for their interest to this work. The author
also thanks Professor Yu.N.Mishurov from Rostov State University,
Rostov-on-Don for giving the useful references.

\bigskip

\end{document}